\documentclass[final,times,5p,twocolumn]{elsarticle}

\usepackage{graphics}
\usepackage{subfigure,graphicx}
\usepackage{epsfig}

\usepackage{amssymb}
\usepackage{amsthm}

\usepackage{amsmath}
\newcommand{\Tr}{\mathrm{Tr}}

\newcommand{\ovl}{\overline}

\journal{Physis Letters B}

\begin{document}

\begin{frontmatter}



\title{Top Quark as a Dark Portal and Neutrino Mass Generation}

\author[label1]{John N. Ng}
\author[label1]{Alejandro de la Puente}
\address[label1]{Theory Group, TRIUMF, Vancouver BC V6T 2A3, Canada}



\begin{abstract}
We present a new model for radiatively generating Majorana active neutrino masses while incorporating a viable dark matter candidate. This is possible by extending the Standard Model with a single Majorana neutrino endowed with a dark parity, a colour electroweak singlet scalar, as well as a colour electroweak triplet scalar. Within this framework, the $up$-type quarks play a special role, serving as a portal for dark matter, and a messenger for neutrino mass generation. We consider three benchmark scenarios where the abundance of dark matter can match the latest experimental results, while generating neutrino masses in the milli-electronvolt range. We show how constraints from lepton flavour violation, in particular the branching fraction of $\mu\to e\gamma$, can place lower bounds on the coupling between our dark matter candidate and top quarks. Furthermore, we show that this coupling can also be constrained using collider data from the Tevatron and the LHC.

\end{abstract}




\end{frontmatter}


\section{Introduction}
\label{sec:Int}

We now have compelling evidence for the existence of three active neutrino species~\cite{Ade:2013zuv}. Radiochemical experiments such as Homestake, Gallex/GNO and SAGE~\cite{Homestake,Gallex,Sage} together with the SuperK and SNO experiments~\cite{SuperK,SNO} have narrowed down the mass patterns to three possibilities: A normal or an inverted hierarchy or almost degenerate masses. Moreover, the absolute scale of neutrino masses remains unknown. Additionally, the last mixing angle, $\theta_{13}$, has been measured by several reactor experiments~\cite{Chooz,Daya,Reno} and the T2K accelerator experiment~\cite{T2K}. This is a breakthrough for the standard picture of neutrino oscillations since it now paves the way towards measuring CP violation in the lepton sector. Furthermore, the evidence for neutrino masses represents one clear motivation for new physics beyond the Standard Model (SM). Within the SM, neutrinos are massless; they can be accommodated in a variety of ways such as incorporating new degrees of freedom and/or new effective interactions. Extending the SM model in this way allows us to be sensitive to new high energy scales. Take for example the Type I seesaw mechanism~\cite{Seesaw}, where the SM is extended with a singlet Majorana fermion that couples to left-handed leptons through the Higgs, as with the charged leptons. This class of models generates viable neutrino masses with a Majorana mass scale $\gtrsim 10^{14}$ GeV and Yukawa interaction of order one. Such a high seesaw scale can arise from Grand Unified models such as SO(10) \cite{GUT}. However, such a high scale for new physics makes the mechanism impossible to test.  A TeV scale Majorana mass is also possible in models such as left-right symmetric models, (for a recent discussion see \cite{LR}). Models with flavour symmetries are also used to explain the neutrino masses (see ~\cite{King:2013eh,Flavor} for recent reviews). Models where neutrino masses are radiatively generated have also been studied. In particular, the simplest model where neutrino masses are induced as one-loop radiative corrections was first introduced in~\cite{Zee}. In this class of models a charged scalar singlet under the SM gauge group couples to left-handed lepton doublets and one is able to generate active neutrino masses of the right order with a charged scalar mass scale as low as a TeV. Neutrino masses may also arise as two loop radiative corrections in extensions of the SM with an additional singlet charged scalar and a doubly charged scalar~\cite{Zee2,Babu,CGNW}. The main motivation for this class of models is that they employ new physics at the TeV scale and hence can be probed at the LHC.

The nature of the neutrino mass matrix can be accessed through data on neutrino oscillations. In the gauge basis the mass matrix can be parametrized in the following way:
\begin{equation}
m_{\alpha\beta}=\sum_{i}m_{i}U_{\alpha i}U^{*}_{\beta i},\label{eq:neutMP}
\end{equation}
where $\alpha,\beta=e,\mu ,\tau$, $i=1,2,3$ and $U_{\alpha,i}$ are the neutrino mixing matrix elements. In general, if neutrinos are Majorana fermions then two new independent degrees of freedom, the Majorana phases, exist and are usually assigned to the $m_{i}$'s. The experimental status on the neutrino oscillation parameters is summarized in~\cite{Dufour}
\begin{eqnarray}
\Delta m^{2}_{21}&=&7.59^{+0.20}_{-0.18}\times 10^{-5}~\text{eV}^{2} \nonumber \\
\sin^{2}\theta_{12}&=&0.312^{+0.017}_{-0.015} \nonumber \\
|\Delta m^{2}_{31}|&=&\left\{
\begin{array}{rl}
2.45\pm0.09\times 10^{-3}~\text{eV}^{2}~\text{Normal~Hierarchy} \\
2.34^{+0.10}_{-0.09}\times 10^{-3}~\text{eV}^{2}~~~\text{Inverted~Hierarchy} \\
\end{array} \right. \nonumber \\
\sin^{2}\theta_{23}&=&0.51\pm0.06.
\end{eqnarray}

Another strong indicator of physics beyond the SM is the ample evidence pointing towards the existence of dark matter~\cite{Bergstrom:2000pn,Bertone:2004pz}. Velocity dispersion and rotation curves of galaxies suggest the existence of non-luminous matter that is not composed by any of the known SM particles. Furthermore, the most recent data from Plank estimates a cold dark matter cosmological parameter $\Omega_{DM}h^{2}=0.1199\pm0.0027$~\cite{Ade:2013zuv} or roughly $26.8\%$ of the universe's total energy. Unfortunately, all experimental evidence for dark matter is due to its gravitational properties and its identity remains unknown to date. One candidate explanation for dark matter is the existence of a weakly interacting massive particle (WIMP). Supersymmetric models are known to provide a natural WIMP candidate, usually the lightest superpartner. The abundance of these particles in the universe is determined by their self-annihilation rate in relation to the expansion of the universe. When the expansion rate dominates over the rate of annihilation, interaction among dark matter particles becomes less efficient and their density becomes a constant or ``freezes out". There is, however, some possible signal regions for WIMP scattering with nuclei in direct detection experiments, most notably the DAMA/LIBRA result~\cite{Dama} and CRESST~\cite{Angloher:2011uu}. Experimental upper limits on the WIMP-nucleon cross section have also been found by various experiments~\cite{Aalseth:2010vx,Agnese:2013rvf,Aprile:2012nq}.

In this work we present a new model for radiatively generating Majorana active neutrino masses while incorporating a viable dark matter candidate. This is possible by extending the SM with a single electroweak singlet Majorana neutrino, $N_R$, to which we also assign an odd parity, referred to as dark parity (DP). We also add a colour electroweak singlet scalar, $\psi$,  and a colour electroweak triplet scalar, $\chi$. These are, respectively, odd and even under DP. Furthermore, all SM fields have even DP. Of the two particles that are odd under DP, we assume $N_R$ to be the lightest. Such an assignment makes $N_R$ a good dark matter candidate since it will be stable, as we engineer DP to be unbroken. It also forbids the usual coupling of $N_R$ to the SM lepton doublet and the Higgs doublet and hence no Dirac mass term is generated. The new colour scalars couple to quark fields, in particular the $up$-type quarks.  In our framework, the $up$-type quarks play two roles: the first one, serving as a messenger for neutrino mass generation. This is possible given the rich structure of the Lagrangian which is used to radiatively generate masses for the left-handed neutrinos via the exchange of the exotic colour scalars at three loops. The second role is as a portal for dark matter, where the relic abundance of dark matter is reproduced through renormalizable interactions between the Majorana neutrino and $up$-type quarks via $\psi$. We study three benchmark scenarios where the abundance of dark matter can match the latest experimental results, while generating neutrino masses in the milli-electronvolt range. Our model is consistent with constraints from lepton flavour violation and collider data from the Tevatron and the LHC. The idea of using a discrete symmetry such as a $Z_2$ parity to forbid a Dirac mass term for the neutrinos and identify $N_R$ as a dark matter candidate was first proposed in \cite{Ma}. Radiative neutrino masses are generated by the use of Higgs triplets or inert doublets. Here we explore a new avenue by making use of colour scalars which allow neutrino masses to be generated at the 3-loop level. Furthermore, the phenomenology at the LHC is richer by virtue that it is very efficient in producing new colour degrees of freedom.


\section{Model}
\label{sec:model}
The model we consider in this study is an extension to the SM that incorporates a dark matter candidate and generates Majorana masses for the active left-handed neutrinos, radiatively and at the three loop level. Within this framework, $N_{R}$ couples to right handed $up$-type quarks through a colour electroweak singlet scalar, $\psi$. Furthermore, we incorporate a coupling between the electroweak lepton doublets and the $up$-type quark doublets through a colour electroweak triplet, $\chi$. The new physics can be parametrized in the following way:
\begin{eqnarray}
\label{eq:lag}
{\cal L}_{BSM}&=&\sum_i y_{\psi}^i\ovl{u^i}P_{L}N^{c}\psi+\sum _{\ell , i} \left\{ \lambda_\ell ^i\left[\ovl{u^i} P_{R}\left(\chi_{1}\nu^{c}_\ell +\chi_{2}\ell^{c} \right)\right.\right. \nonumber \\
&+&\left.\left.\ovl{d^i} P_{R}\left(\chi_{3}\ell^{c} -\chi_{2}\nu^{c}_\ell \right)\right]\right\}+\text{hc},
\end{eqnarray}
where $l=e,\mu,\tau$ and $i=1,2,3$ is the quark family index. The coupling $y^{i}_{\psi}$ denotes the strength of the interaction between $N_{R}$ and $u^{i}_{R}$ via $\psi$, while $\lambda^{i}_{l}$ the strength between the quark doublets $(u^{i},d^{i})_{L}$ and $(\nu,l)_{L}$ via $\chi$. Throughout this work, we make use of $P_{R/L}=\frac{1\pm \gamma_{5}}{2}$. Furthermore, unless otherwise stated, we work in the charged fermion mass basis.

 Under the SM gauge group $SU(3)_c\times SU(2)_{W}\times U(1)_Y$, $\psi$ transforms as a ${\bf (3,1,2/3)}$ and we write the field $\chi$  as
\begin{equation}
\chi=\begin{pmatrix}
\chi_{2}/\sqrt{2} & \chi_{1} \\
\chi_{3} & -\chi_{2}/\sqrt{2}
\end{pmatrix}
\end{equation}
which transforms as a ${\bf (3,3,-1/3)}$. These assignments yield electric charges of $Q=2/3,-1/3,-4/3$ for $\chi_{1},\chi_{2}$ and $\chi_{3}$ respectively.

The gauge covariant derivatives for the scalars are given by
\begin{equation}
{\cal L}_{kin}=(D_{\mu}\psi)^{\dagger}(D^{\mu}\psi)+\Tr(D_{\mu}\chi)^{\dagger}(D^{\mu}\chi),
\end{equation}
where
\begin{eqnarray}
D_{\mu}\psi&=&\left(\partial_{\mu}-ig_{s}G^{a}_{\mu}\lambda^{a}-ig'(\frac{2}{3})B_{\mu}\right)\psi \nonumber \\
D_{\mu}\chi&=&\partial_{\mu}-ig_{s}G^{a}_{\mu}\lambda^{a}\chi-\frac{ig}{2}[W^{i}_{\mu}\sigma^{i},\chi]-ig'(\frac{-1}{3})B_{\mu}\chi. \nonumber \\
\end{eqnarray}
The implicit sums are over the generators $\lambda^a$ of SU(3), $a=1,...8$,  and the generators $\sigma^i$ of SU(2), $i=1,2,3$.

Within this framework a $N_{R}$ Majorana mass term,  $\frac{1}{2}\, M_{N_{R}}\,\bar{N^c_{R}}\,N_R$, can be added. This term is even under DP, and we treat $M_{N_{R}}$ as a free parameter. We further assume that $M_{N_{R}} < m_\psi$ which makes $N_{R}$ a suitable dark matter candidate\footnote{We have assumed that $m_{\psi}>M_{N_{R}}$, such that $M_{N_{R}}$ is the lightest stable particle under the DP. We may also have $M_{N_{R}}>m_{\psi}$. In the latter case, our model will be one with a strongly interacting dark matter candidate, an analysis that is beyond the scope of this paper.}.

The gauge and  $Z_2$ invariant potential is given by
\begin{eqnarray}
\label{eq:spot}
V(H,\psi,\chi)= -\mu^2 H^{\dagger} H +\frac{\lambda}{4!}(H^{\dagger} H)^2 +m_\chi^2 \Tr (\chi^{\dagger}\chi) \nonumber \\
+ \lambda_\chi (\Tr (\chi^{\dagger} \chi))^2 + m_\psi ^2 \psi^{\dagger}\psi + \lambda_\psi (\psi^{\dagger}\psi)^2+\kappa_1 H^{\dagger} H \Tr (\chi^{\dagger} \chi)  \nonumber \\
+\kappa_2 H^{\dagger} \chi^{\dagger} \chi H + \kappa_3 H^{\dagger} H \psi^{\dagger} \psi +\rho \Tr (\chi^{\dagger} \chi)\psi^{\dagger} \psi
\end{eqnarray}
where $H$ is the SM Higgs field. In order not to have a colour breaking vacuum we take $m_\chi^2,m_\psi^2$ to be positive. Since all the new physics terms we have incorporated are of dimension four, the theory remains renormalizable. Of particular interest to us is the last term of Equation (\ref{eq:spot}), since this coupling would play a role in the radiative generation of the active neutrino masses. In addition, the DP remains exact even after electroweak symmetry breaking.

\section{Dark Matter}\label{sec:darkmatter}

As mentioned in the previous section, the unbroken  $Z_{2}$ symmetry stabilizes $N_R$. Due to the interaction introduced in Equation (1), the mechanism that leads to a reduction in the relic abundance of $N_{R}$ is via $t$-channel annihilation into right-handed top and charm quarks through the exchange of the colour electroweak singlet scalar, $\psi$. In this work we consider three benchmark points which depict three important regions of parameter space: $M_{N_{R}}=80,150,450$ GeV.

\begin{figure}[ht]\centering
\includegraphics[width=3.5in]{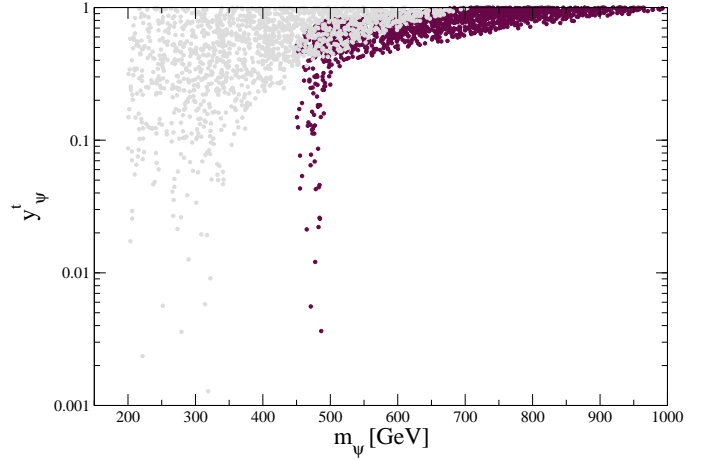}
\caption{The normalized relic abundance in the $y^{t}_{\psi}-m_{\psi}$ plane. The grey region corresponds to the region of parameter space consistent with a Majorana neutrino with mass $M_{N_{R}}=150$ GeV contributing $75-100\%$ of the dark matter relic abundance. The region in maroon corresponds to $M_{N_{R}}=450$ GeV.   \label{fig:RelicY1}}
\bigskip
\end{figure}

\begin{figure}[ht]\centering
\includegraphics[width=3.5in]{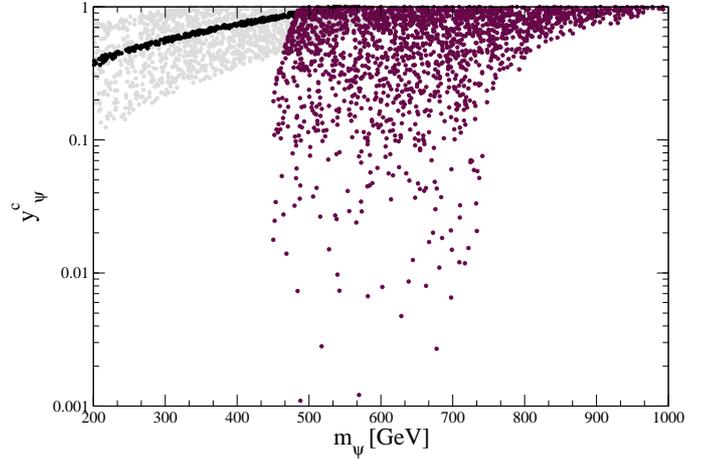}
\caption{The normalized relic abundance in the $y^{c}_{\psi}-m_{\psi}$ plane. The black region corresponds to the region of parameter space consistent with a Majorana neutrino with mass $M_{N_{R}}=80$ GeV contributing $75-100\%$ of the dark matter relic abundance. The region in grey and maroon correspond to $M_{N_{R}}=150,450$ GeV respectively.  \label{fig:RelicY2}}
\end{figure}

The evolution of the comoving particle density is given by the Boltzmann equation
\begin{equation}
\frac{\dot{n}}{n_{eq}}=\Gamma\cdot\left(\frac{n^{2}}{n^{2}_{eq}}-1\right)-3H\frac{n}{n_{eq}}
\end{equation}
where $n$ is the particle density at time $t$ and $n_{eq}$ is the density at equilibrium, $H$ is the Hubble expansion rate and $\Gamma$ parametrizes the interaction rate, $\Gamma=\left<\sigma v\right>n_{eq}$, where $\left<\sigma v\right>$ denotes the thermally average annihilation cross section. By solving numerically the above equation one can find the temperature at which particles depart from equilibrium and freeze out. This temperature is given by
\begin{equation}
x_{FO}\equiv\frac{m}{T_{FO}}\approx\log\left(0.038g\frac{mM_{Pl}\left<\sigma v\right>}{g^{1/2}_{*}x^{1/2}_{FO}}\right),
\end{equation}
where $g$ denotes the number of degrees of freedom of the particle under consideration and $g_{*}$ the number of relativistic degrees of freedom at the freeze out temperature. The present day relic abundance is then given by
\begin{equation}
\Omega_{DM}h^{2}\approx\frac{1.07\times10^{9}~\text{GeV}^{-1}}{Jg^{1/2}_{*}M_{Pl}},
\end{equation}
where
\begin{equation}
J\equiv\int^{\infty}_{x_{FO}}\frac{\left<\sigma v\right>}{x^{2}}dx.
\end{equation}
The thermalized cross section at temperature $T$ can be calculated from the annihilation cross section of our dark matter candidate, $N_{R}$. The thermalized cross section is given by
\begin{equation}
\left<\sigma_{N_{R}N_{R}} v\right>=\int^{\infty}_{4M^{2}_{N_{R}}}ds \frac{(s-4M^{2}_{N_{R}})s^{1/2}K_{1}\left(s^{1/2}/T\right)}{8M_{N_{R}}TK_{2}^{2}(M_{N_{R}}/T)}\sigma(s),
\end{equation}
where $\sigma(s)$ is the annihilation cross section as a function of the center of mass energy squared of the interaction, and $K_{1}(z)$, $K_{2}(z)$ are Modified Bessel function of the first and second kind respectively. We calculated the relic abundance using the latest version of MicOMEGAs~\cite{Belanger:2010gh}
and the model files were generated with the latest version of FeynRules~\cite{Feynr}. We carried out a scan over three parameters, $y_{\psi}^{t,c}$, and $m_\psi$, for the three benchmark points. The results are shown in Figures~\ref{fig:RelicY1} and~\ref{fig:RelicY2}.

\vspace{1.0 cm}
\vspace{1.0 cm}
The dependence of the relic abundance on $y^{t}_{\psi}$ and $m_{\psi}$ is shown in Figure~\ref{fig:RelicY1}. The grey region denotes the parameter space consistent with a relic with mass $M_{N_{R}}=150$ GeV contributing $75-100\%$ of the dark matter relic abundance, while the maroon region is for a relic with mass $M_{N_{R}}=450$ GeV. In Figure~\ref{fig:RelicY2} we show the dependence of the relic abundance as a function of $y^{c}_{\psi}$ and $m_{\psi}$. The grey and maroon regions correspond to a relic with mass of $150$ and $450$ GeV respectively. The scattered behaviour of the grey and maroon regions in Figures~\ref{fig:RelicY1} and~\ref{fig:RelicY2} is due to the fact that a combination of annihilation channels are open: $c\bar{c}$ and $t\bar{c}/c\bar{t}$ for a $150$ GeV Majorana neutrino and $c\bar{c}$, $t\bar{t}$ and $t\bar{c}/c\bar{t}$ for a $450$ GeV Majorana neutrino. This is not the case for a relic with $M_{N_{R}}=80$ GeV, where the $c\bar{c}$ annihilation channel is the only one open. Here one finds that the relic abundance depends only on $y^{c}_{\psi}$ and $m_{\psi}$. For an $80$ GeV Majorana neutrino, the region consistent with $75-100\%$ of the relic abundance is depicted by the black region in Figure~\ref{fig:RelicY2}.

\section{Radiative Neutrino Mass generation}\label{sec:neutmass}

\begin{figure}[ht]\centering
\includegraphics[width=3.5in]{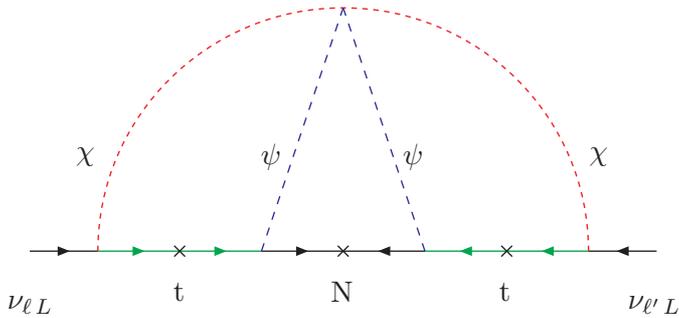}
\caption{3-loop generation of a Majorana mass for active neutrinos from the t-quark. The crosses on the fermion lines indicate mass insertions. Similar diagrams from the c-quark will also play a role although it gives smaller contribution. \label{fig:numass_plot}}
\end{figure}

The conserved DP allows us to identify $N_{R}$ as a candidate for dark matter and it also forbids Dirac neutrino mass terms for the active neutrinos, $\nu_{i}$. Therefore, the usual seesaw mechanism is not operative in this model. However, the Lagrangian of Equation (\ref{eq:lag}) has enough structure to radiatively generate masses for $\nu_{i}$ via the exchange of the exotic colour scalars. In particular, it has the novel feature of using the $t_R$ and $c_{R}$ quarks as a portal to communicate with the dark sector and as messengers for the neutrinos. Within this framework, the lowest order diagram for neutrino mass generation is at 3-loops. The diagram is due to exchanges of both $\psi$ and $\chi$ fields. This is depicted in Figure \ref{fig:numass_plot} which gives the $\ell,\ell^\prime$ element of the active neutrino mass matrix $M_\nu$.

 This mechanism yields finite contributions to all the elements of $M_{\nu}$ and it is best seen using the mass insertion technique. The $\ell \ell^\prime$ element of the
 active neutrino mass matrix is given by
\begin{equation}
\label{eq:numass}
(M_\nu)_{\ell \ell^\prime}=\sum_{i,j}K^{ij}\lambda_\ell^{i} \lambda_{\ell^\prime}^{j}
\end{equation}
where $i,j=u,c,t$ and $K^{ij}$ which controls the scale of neutrino masses is given by
\small
\begin{eqnarray}
 K^{ij}&=&\frac{y_\psi^{i} y^{j}_\psi \rho}{(16\pi^2)^3} \frac{m_i m_j\, M_{N_{R}}}{(m_\chi^2-m_i^2)(m^2_{\chi}-m_j^{2})}I(m_\chi ^2,m_\psi^{2}),\nonumber \\
 I&=& \int_0^{\infty} du \, \frac{u}{u+M_{N_{R}}^{2}} \nonumber \\
 &\cdot&\left[\int_0^{1}dx\ln\left(\frac{m_\chi^{2}(1-x)+ m_\psi^{2}x +u x(1-x)}{m_i ^{2}(1-x)+ m_\psi^{2}x +u x(1-x)}\right)\right]^2.
 \end{eqnarray}
 \normalsize
From the above equation we see that the $u$-quark yields a negligible contribution to the neutrino masses and we can concentrate on the $t$ and $c$ quarks. Furthermore, if only one type of quark is involved in the neutrino mass generation, then Equation (\ref{eq:numass}) gives rise to two massless active neutrinos, excluded by experimental data. Therefore, at least two quark families must come into play. To simplify the model we assume that the top quark gives the main contribution and also demand that $\lambda_{e,\mu}^c <<\lambda^c_\tau$, such that the $c$-quark contribution only modifies the $3,3$ element of $M_\nu$. These requirements are sufficient to lift the degeneracy of two massless neutrinos.
\begin{figure}[ht]\centering
\bigskip
\bigskip
\includegraphics[width=3.5in]{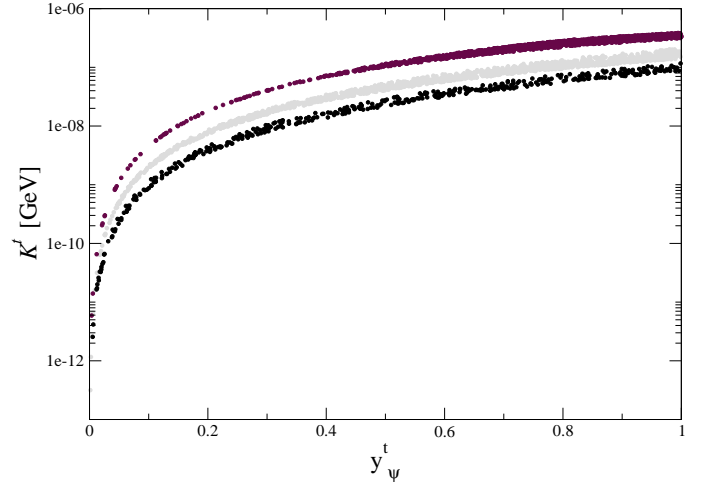}
\caption{$K^{t,t}$ factor as a function of the $N_{R}-t_{R}$ coupling, $y^{t}_{\psi}$. The region in black corresponds to a Majorana neutrino with $M_{N_{R}}=80$ GeV while the grey and maroon regions correspond to Majorana neutrino masses of $150$ and $450$ GeV respectively.
\label{fig:Kfac}}
\end{figure}

Using this framework for neutrino mass generation we analyzed the parameter space consistent with 75-100$\%$ of the dark matter relic abundance, and calculated the $K^{ij}$ factors. In Figure~\ref{fig:Kfac} we show the $K^{t,t}$ factor as a function of $y^{t}_{\psi}$. The black region corresponds to a Majorana neutrino with $M_{N_{R}}=80$ GeV and the grey and maroon regions correspond to Majorana neutrino masses of $150$ and $450$ GeV respectively. We use a colour electroweak triplet with mass $m_{\chi}=1$ TeV and a scalar potential coupling between $\chi$ and $\psi$ of $\rho=0.1$. The bulk of the neutrino mass is due to $K^{t,t}$ since $K^{t,t}\gg K^{c,c}$. The $K^{t,t}$ parameter ranges from $\sim$one meV to 100 eV for parameter points responsible for 75-100$\%$ of the dark matter relic abundance. This range of $K^{t,t}$ values can naturally provide this model with a milli-electronvolt active neutrino mass. It is easy to see why the neutrino masses are naturally small. Let us consider the t-quark contribution. The 3-loop suppression yields a factor of $10^{-7}$. Since the LHC has not seen any new colour states we can assume that $m_\chi > 1~\mathrm{TeV}$. A further suppression comes from $(\frac{m_t}{m_\chi})^2\sim 10^{-2}$. For $M_{N_{R}}=100$ GeV, the factor $(\frac{M_{N_R}}{m_\chi})^2$ gives another $10^{-2}$ suppression. Therefore, sub-eV active neutrinos are natural in this model
and no fine tuning of $y_{\psi}^{t,c}$ or $\rho$ is required.

\section{$\mu\to e\gamma$}\label{sec:mutoeg}

\begin{figure}[ht]\centering
\includegraphics[width=3.5in]{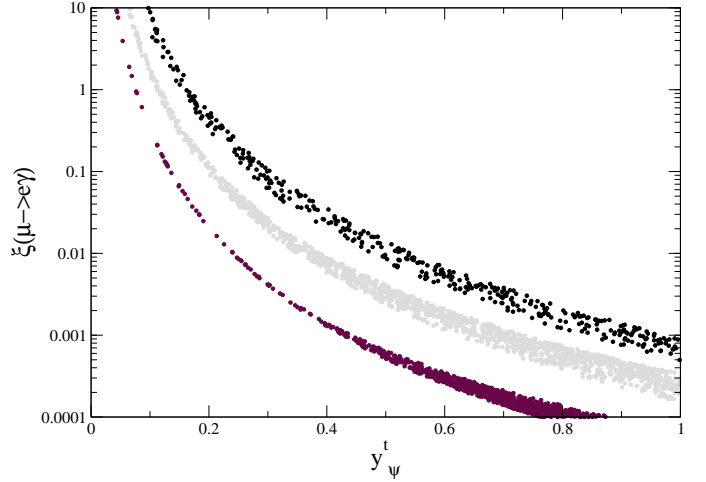}
\caption{Lower limit on the branching fraction normalized to the experimental upper bound as a function of $y^{t}_{\psi}$ using the best fit values for $m_{e\mu}$ using an electroweak triplet scalar mass, $m_{\psi}=1$ TeV. The black region corresponds to $M_{N_{R}}=80$ GeV while the grey and maroon regions correspond to $M_{N_{R}}=150,450$ GeV respectively.  \label{fig:MuEG_LB}}
\bigskip
\bigskip
\end{figure}

\begin{figure}[ht]\centering
\includegraphics[width=3.5in]{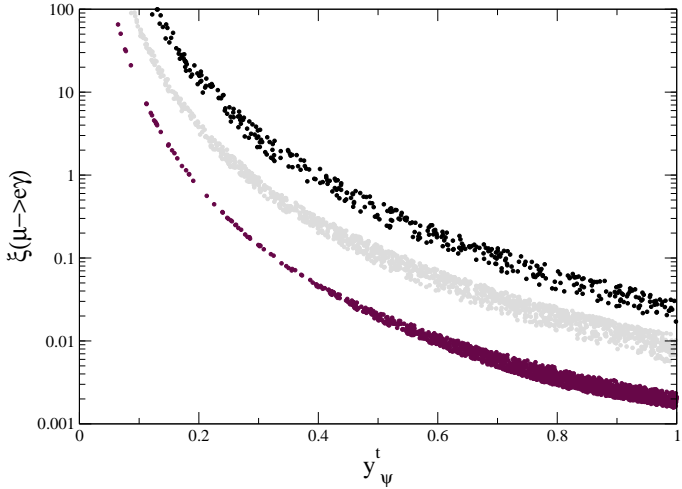}
\caption{Upper limit on the branching fraction normalized to the experimental upper bound as a function of $y^{t}_{\psi}$ using the best fit values for $m_{e\mu}$ using an electroweak triplet scalar mass, $m_{\psi}=1$ TeV. The black region corresponds to $M_{N_{R}}=80$ GeV while the grey and maroon regions correspond to $M_{N_{R}}=150,450$ GeV respectively. \label{fig:MuEG_UB}}
\end{figure}

From Equation~\ref{eq:lag} one can see that the colour electroweak triplet scalar states will give rise to lepton flavour violating decays. In particular, the decay $\mu\to e\gamma$ can be used to place a lower bound on the $y^{t}_{\psi}$ coupling. In our framework, the branching fraction of $\mu\to e \gamma$ is given by
\begin{equation}
Br(\mu\to e \gamma) =1.8\left( \frac{ \mathrm{TeV}}{m_\chi}\right)^4 \times 10^{-6}|\lambda_\mu ^t\lambda_e ^t +\lambda_\mu ^c\lambda_e ^c|^2.
\end{equation}
Given that $K^{t,t}\gg K^{c,c}$, we see that we have no sensitivity to $\lambda^{c}_{\mu}\lambda^{c}_{e}$ in the definition of $M_{\nu}$. In this work we have maximized the contribution to the branching fraction in the limit where $\lambda^{c}_{\mu}\lambda^{c}_{e}\sim\lambda^{t}_{\mu}\lambda^{t}_{e}$. We then extract the value of $\lambda^{t}_{\mu}\lambda^{t}_{e}$ using the results from Figure~\ref{fig:Kfac} together with the latest values of $m_{e\mu}$~\cite{Bertuzzo:2013ew} and the current experimental upper bound on $Br(\mu\to e\gamma)\le2.4\times10^{-12}$~\cite{Adam:2011ch}. In the analysis, we have used the best fit range for $m_{e\mu}$ assuming a normal hierarchy of active neutrino masses, $|m_{e\mu}|=1.5-8.8$ meV~\cite{Bertuzzo:2013ew}. We have also fixed the colour electroweak triplet mass to $m_{\chi}=1$ TeV. The branching fraction can then be written in the following way:
\begin{equation}
Br(\mu\to e\gamma)=7.2\times 10^{-6}\left(\frac{m_{e\mu}}{K^{t,t}}\right)^{2}
\end{equation}
Our results are shown in Figure~\ref{fig:MuEG_LB}, where we plot the normalized branching fraction, $\xi(\mu\to e\gamma)=Br(\mu\to e\gamma)/Br(\mu\to e\gamma)_{exp}$, as a function of $y^{t}_{\psi}$ using the lower limit on $m_{e\mu}$; and in Figure~\ref{fig:MuEG_UB}  using the upper limit on $m_{e\mu}$. The black region corresponds to $M_{N_{R}}=80$ GeV while the grey and maroon regions to $M_{N_{R}}=150,450$ GeV respectively. We see that the lower bound on $y^{t}_{\psi}$ increases with decreasing $M_{N_{R}}$. This behaviour is due to the fact that the branching fraction is proportional to $M^{2}_{N_{R}}$ while it is inversely proportional to $(y^{t}_{\psi})^{4}$. In particular, we find an upper bound of $y^{t}_{\psi}\lesssim 0.3-0.4$ for $M_{N_{R}}=80$ GeV and $y^{t}_{\psi}\lesssim0.2-0.3,0.18-0.2$ for $M_{N_{R}}=150,450$ GeV.
\vspace{1.0 cm}
\vspace{1.0 cm}

An important fact to note is that the constraints placed on $y^{t}_{\psi}$ using the current experimental bound on $Br(\mu\to e\gamma)$ are not at all sensitive to the mass of the colour electroweak singlet scalar. This scalar plays an important role in mediating the annihilation of the Majorana neutrinos. As we will see below, bounds on the mass of this scalar as well as upper bounds on the $y^{t}_{\psi}$ can be obtained using collider data.

\section{Collider constraints}\label{sec:collider}

This model is also highly constrained by data from high energy colliders such as the Tevatron and the LHC. In particular, our model yields two very distinct signatures for which very stringent bounds exist. We used Madgraph 5~\cite{Alwall:2011uj} to calculate the parton-level signal prediction and implemented the initial and final state radiation using Pythia~\cite{Sjostrand:2006za}. Our signal acceptances were are calculated with the PGS detector simulation implementing the cuts in the corresponding LHC and Tevatron analyses.
\begin{figure}[ht]\centering
\subfigure{\includegraphics[width=3.0in]{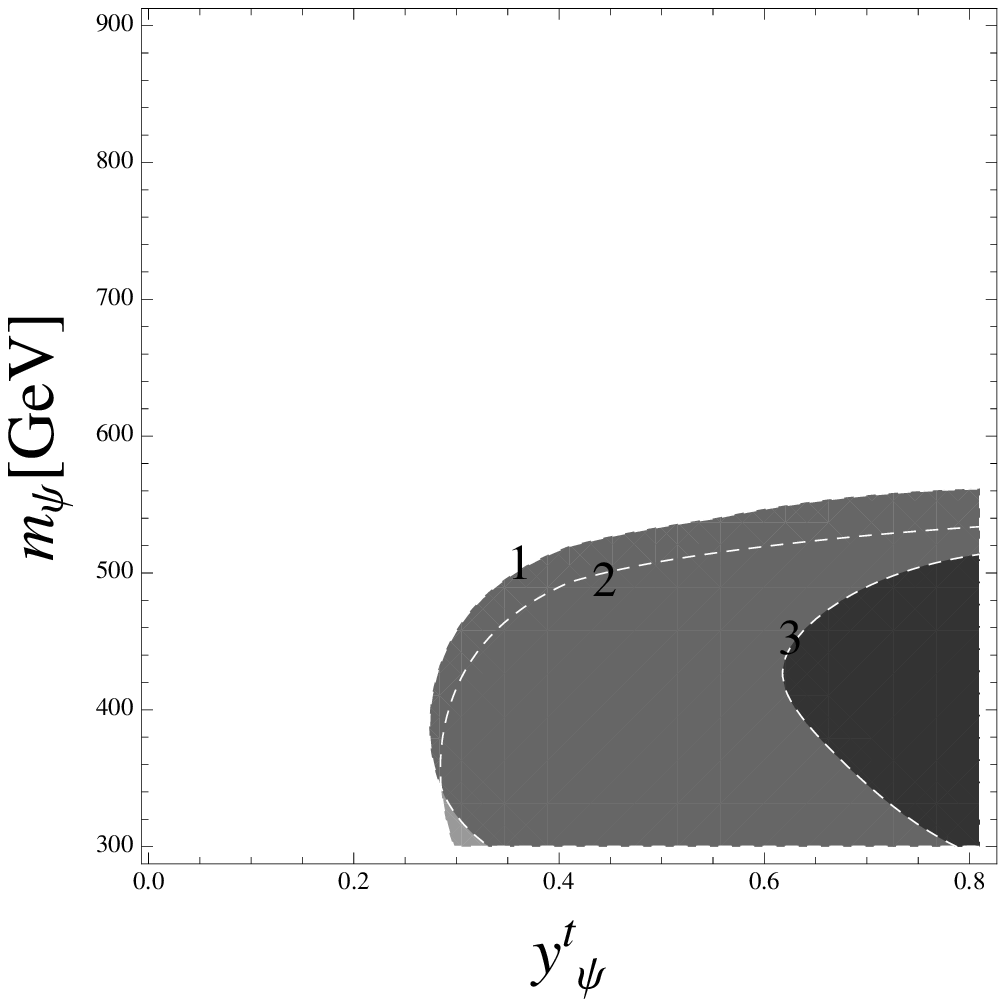}}
\bigskip
\subfigure{\includegraphics[width=3.0in]{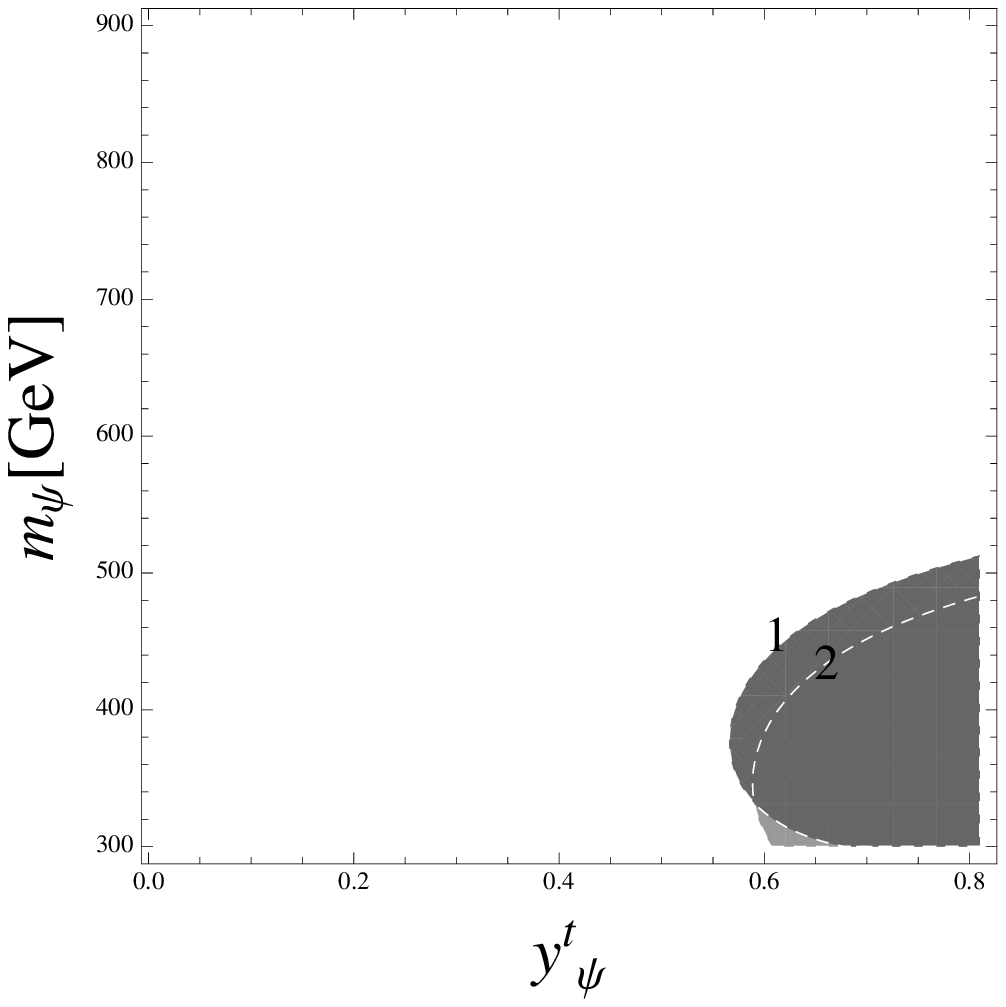}}
\caption{Collider constraints for an 80 GeV Majorana neutrino from the four experimental observables mentioned at the beginning of this section. On the top we show the excluded region in the $y^{t}_{\psi}-m_{\psi}$ plane for $y^{c}_{\psi}=0.25$. The region labeled 1 corresponds to regions of parameter space excluded by the CMS observable with MET~$>200$ GeV, while the  regions labeled 2 and 3 correspond to MET~$>150$ and $>300$ GeV respectively. The plot on the bottom corresponds to $y^{c}_{\psi}=0.5$.
 \label{fig:ExMnr80}}
\end{figure}

One constraint is due to dijet plus missing energy (MET) searches at the Tevatron. The latest bounds on this process were carried out by the CDF collaboration using $p\bar{p}$ collisions at a center of mass energy of $\sqrt{s}=1.96$ TeV and $2.$ fb$^{-1}$ of integrated luminosity~\cite{Aaltonen:2009xp}. Within our framework, two channels can lead to this final state. The first one is $t\bar{t}$ production followed by a three body decay of the top quark into two Majorana neutrinos and a charm quark, $t\to N_{R}N_{R}c$. This channel is open as long as $N_{R}$ has a mass below $\sim86$ GeV. The second channel is through pair production of two colour electroweak singlets, followed by the decay $\psi\to N_{R}c$. These two channels are sensitive to $y_{\psi}^{t,c}$ and $m_{\psi}$. In order to generate exclusions on all three parameters of our model we implemented the experimental sample with tight kinematic thresholds of MET~$>100$ GeV and $H_{T}>225$ GeV, where $H_{T}$ denotes the scalar sum of the two jet transverse energies:
\begin{equation}
H_{T}=E_{T}(\text{jet}_{1})+E_{T}(\text{jet}_{2})
\end{equation}

The second constraint is due to top squark pair production in $pp$ collisions with a center of mass energy of $\sqrt{s}=8$ TeV and $19.5$ fb$^{-1}$ of integrated luminosity. We used the results obtained with the Compact Muon Solenoid (CMS) detector at the LHC. This search looks for decays of a stop squark into a top quark and a neutralino~\cite{CMS:2011zra}. Top squarks are the scalar partners of the top quark in supersymmetric extensions of the SM such as the Minimal Supersymmetric Standard Model (MSSM), and the neutralino is a linear combination of the fermionic partners of the neutral gauge bosons and the two neutral Higgs bosons. Within our framework, the colour electroweak singlet, $\psi$, has the same gauge quantum numbers as the top squark but additional decay modes, in particular $\psi\to N_{R} c$. We apply the CMS constraint using their cut based analysis for three different MET cuts: $>150,200,300$ GeV.

In Figure~\ref{fig:ExMnr80}, we show the parameter regions excluded for an 80 GeV Majorana neutrino from the four experimental observables mentioned at the beginning of this section. On the top, we plot the excluded region in the $y^{t}_{\psi}-m_{\psi}$ plane for $y^{c}_{\psi}=0.25$. The region labeled 1 corresponds to regions of parameter space excluded by the CMS observable with MET~$>200$ GeV, while the  regions labeled 2 and 3 correspond to MET~$>150$ and $>300$ GeV respectively. The plot at the bottom corresponds to a value of $y^{c}_{\psi}=0.5$. For this value of $y^{c}_{\psi}$ the excluded region is smaller since the branching fraction of $\psi\to N_{R}t$ is reduced, and thus, the CMS analysis is less sensitive to our model. From the plots in Figure~\ref{fig:ExMnr80} we also see that no region is excluded by the CDF experiment for $m_{\psi}>300$ GeV. This is not true for masses below $300$ GeV, where the CDF experiment rules out the entire model for $M_{N_{R}}=80$ GeV.

\begin{figure}[ht]\centering
\subfigure{\includegraphics[width=3.0in]{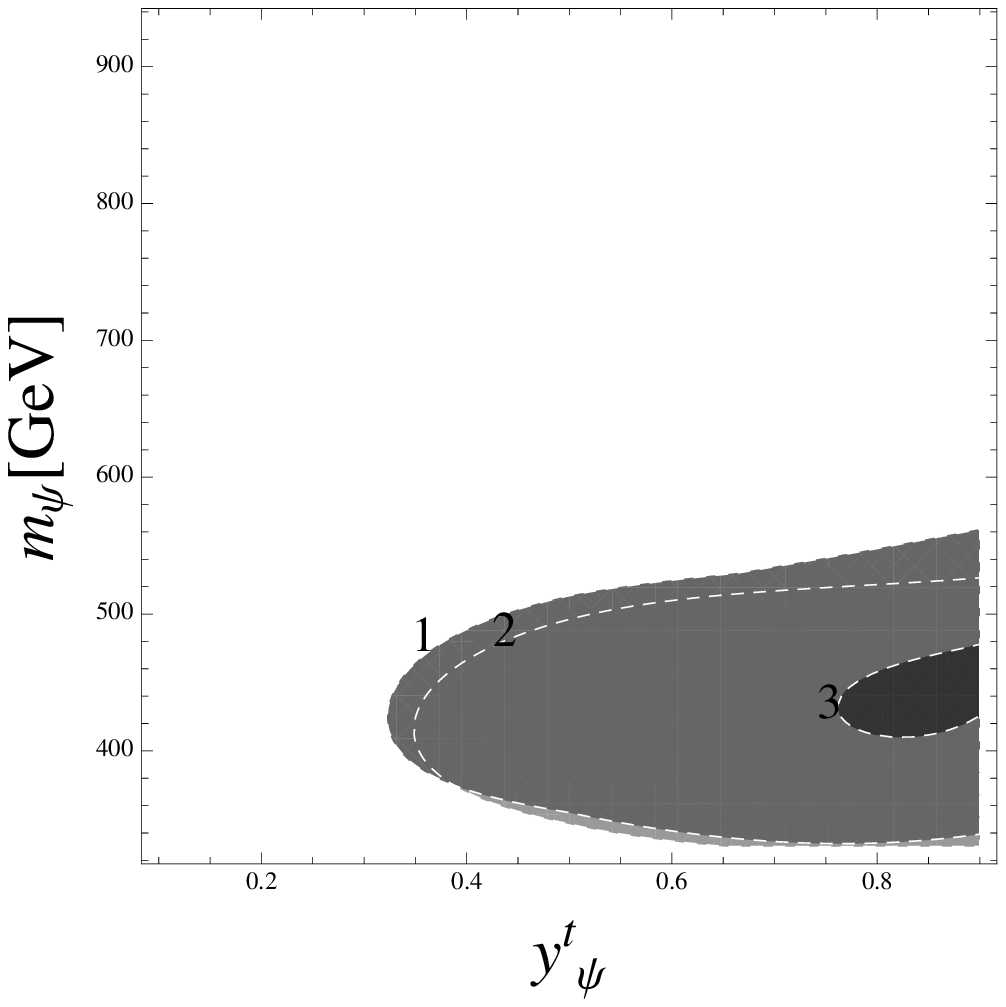}}
\bigskip
\subfigure{\includegraphics[width=3.0in]{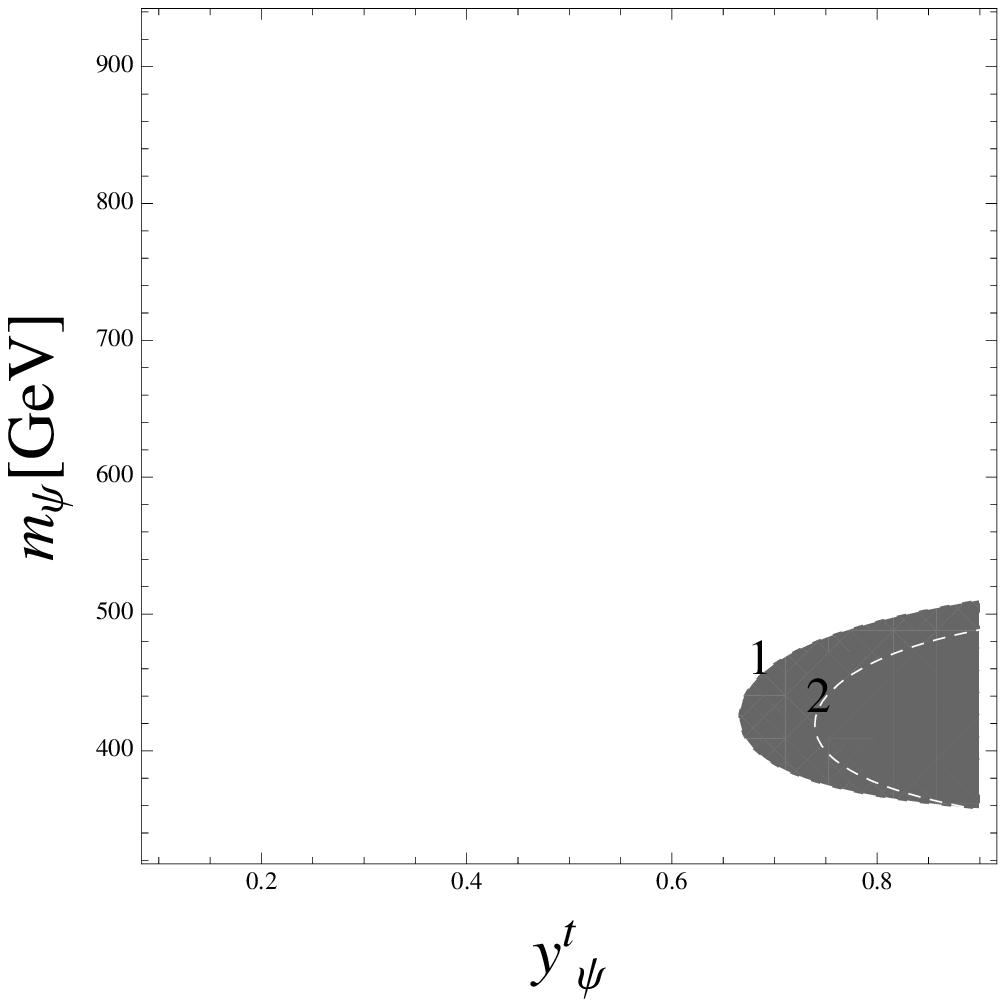}}
\caption{Collider constraints for a 150 GeV Majorana neutrino from the four experimental observables mentioned at the beginning of this section. On the top we show the excluded region in the $y^{t}_{\psi}-m_{\psi}$ plane for $y^{c}_{\psi}=0.25$. The region labeled 1 corresponds to regions of parameter space excluded by the CMS observable with MET~$>200$ GeV, while the  regions labeled 2 and 3 correspond to MET~$>150$ and $>300$ GeV respectively. The plot on the bottom corresponds to $y^{c}_{\psi}=0.5$.
 \label{fig:ExMnr150}}
\end{figure}
In Figure~\ref{fig:ExMnr150}, we show the regions of parameter space excluded for a 150 GeV Majorana neutrino. The plot on the top corresponds to $y^{c}_{\psi}=0.25$ while the plot on the bottom to $y^{c}_{\psi}=0.5$. For this benchmark point, the CDF observable is sensitive to regions where $m_{\psi}$ lies above 200 GeV but it is not able to exclude any of that region of parameter space. Therefore, the only relevant observable is the CMS analysis, which is able to exclude a region of parameter space where $320\lesssim m_{\psi}\lesssim 550$ for $y^{t}_{\psi}\gtrsim 0.4$ and $y^{c}_{\psi}=0.25$. Again, the excluded region is significantly smaller for larger values of $y^{c}_{\psi}$, since the branching fraction of $\psi\to N_{R}t$ is suppressed.

The above collider constraints were also applied to a Majorana neutrino with $M_{N_{R}}=450$ GeV. We found that these constraints were not strong enough to rule out any of the parameter space consistent with $75-100\%$ of the dark matter relic abundance. Furthermore, we found that for Majorana neutrinos with masses below $20$ GeV, the CDF data on dijet+MET was enough to exclude it as a viable dark matter candidate.


\section{Discussion}\label{sec:conclusions}

In this study, we have investigated the possibility of extending the Standard Model with an electroweak singlet Majorana neutrino, stabilized by a new $Z_{2}$ symmetry, to explain the abundance of the dark matter in the universe. In this model, we coupled the dark matter candidate to $up$-type quarks via a new colour electroweak singlet scalar. Throughout the study we considered three benchmark scenarios: $M_{N_{R}}=80,150,450$ GeV. We found that the main annihilation channels were into right-handed top and charm quarks depending on the Majorana neutrino mass. Furthermore, we found that when all annihilation channels were open, we were able to generate $75-100\%$ of the dark  matter relic abundance with a wide range of couplings, $y^{t,c}_{\psi}$, and scalar masses, $m_{\psi}$. This however was not the case for $M_{N_{R}}=80$ GeV, where the only available annihilation channel was into charm quarks. In this case we found a very clear dependence of the coupling $y^{c}_{\psi}$ on $m_{\psi}$.

We have also investigated the possibility of radiatively generating Majorana masses for the active neutrinos of the Standard Model by incorporating a colour electroweak triplet scalar in addition to the colour electroweak singlet scalar. This setup allowed us to generate active neutrino masses at three loops. We found that the neutrino mass was mostly sensitive to the $y^{t}_{\psi}$ coupling, and that for points consistent with $75-100\%$ of the dark matter relic abundance, neutrino masses in the meV to 100 eV range are natural, with data favouring the lower values.

We have considered two types of constraints. The first one arising from the lepton flavour violating decay, $\mu\to e\gamma$. We found that the current experimental bound on the branching fraction placed lower bounds on the coupling $y^{t}_{\psi}$ independent on the colour electroweak singlet mass, $m_{\psi}$. This lower bound was also higher for lighter Majorana neutrinos. The second constraint was due to two different collider searches. We found that these constraints place upper bounds on the coupling $y^{t}_{\psi}$. These constraints were also dependent on $m_{\psi}$ and $y^{c}_{\psi}$; the latter responsible for the size of the excluded region, since this coupling modifies the branching fraction of $\psi\to N_{R}t$.

Our framework offers an attractive avenue that naturally generates small active neutrino masses while providing a large range of masses for a viable dark matter candidate. The model we presented here is a minimal one as only couplings to $t$ and $c$ quarks are employed. The model also predicts new colour degrees of freedom which lie below the TeV scale, and are now being probed at the LHC. Further signatures at the LHC, such as rare top quarks decays, monotop production and effects on the LHC Higgs signals, will be reported elsewhere.

\section*{Acknowledgements}

ADP would like to thank Jorge de Blas Mateo and Travis Martin for useful discussions and essential feedback regarding the progress of this work. This work is supported in parts by the National Science and Engineering Council of Canada.




\bibliographystyle{elsarticle-num}
\bibliography{<your-bib-database>}



\end{document}